\documentclass[copyright]{eptcs}
\usepackage{ifxetex}
\ifxetex \else \pdfoutput=1 \fi  
\input{./tex/preamble}%
\renewcommand{\include}{\input}
\begin{document}
\maketitle

\begin{abstract}
This paper proposes a language for describing reactive synthesis problems that integrates imperative and declarative elements.
The semantics is defined in terms of two-player turn-based infinite games with full information.
Currently, synthesis tools accept linear temporal logic (LTL) as input, but this description is less structured and does not facilitate the expression of sequential constraints.
This motivates the use of a structured programming language to specify synthesis problems.
Transition systems and guarded commands serve as imperative constructs, expressed in a syntax based on that of the modeling language \textsc{Promela}.
The syntax allows defining which player controls data and control flow, and separating a program into assumptions and guarantees.
These notions are necessary for input to game solvers.
The integration of imperative and declarative paradigms allows using the paradigm that is most appropriate for expressing each requirement.
The declarative part is expressed in the LTL fragment of generalized reactivity(1), which admits efficient synthesis algorithms, extended with past LTL.
The implementation translates \textsc{Promela} to input for the \textsc{Slugs} synthesizer and is written in \textsc{Python}.
The AMBA AHB bus case study is revisited and synthesized efficiently, identifying the need to reorder binary decision diagrams during strategy construction, in order to prevent the exponential blowup observed in previous work.
\end{abstract}

\section{Introduction}

Over the past three decades, system formal verification has aided design and become practical for industrial application.
In the past decade, synthesis of systems from specifications has seen significant development \cite{Kupferman12sofsem,Walukiewicz04lics}, partially owing to the discovery of temporal logic fragments that admit efficient synthesis algorithms \cite{Piterman06,Bloem12jcss,Ehlers11nfm,Alur04tocl}.
Applications range from protocol synthesis for hardware circuits \cite{Bloem12jcss}, to correct-by-construction controllers for hybrid systems \cite{KressGazit09tro,Kloetzer08tac,Wongpiromsarn13us}.

Many languages and tools have been developed for modeling and model checking systems.
Unlike verification using model checking, the tools for synthesis have been developed much more recently.
One reason is that centralized synthesis from linear temporal logic (LTL) \cite{Pnueli77} has doubly exponential complexity in the length of the specification formula \cite{Rosner92}, a result that did not encourage further development initially.

Currently, LTL is the language used for describing specifications as input to synthesis tools.
There are many benefits in using a logic for synthesis tasks, its declarative nature being a major one, because it allows expressing individual requirements separately, and in a precise way.
It also makes explicit the implicit conventions present in programming languages \cite{Lamport94tpls}.
Another aspect of synthesis problems that makes declarative descriptions appropriate is that we want to describe as large a set of possible designs as possible, in order to avoid overconstraining the search space.

However, not all specifications are best described declaratively.
There exist synthesis problems whose description involves graph-like structures that are cumbersome for humans to write in logic.
Robotics problems typically involve graph constraints that originate from possible physical configurations.
For example, considering a wheeled robot, its physical motion is modeled by possible transitions that avoid collisions with other objects, whereas an objective to patrol between two locations can more appropriately be described with a temporal logic formula.
Properties that specify sequential behavior also lead to graph-like structures, and require use of auxiliary variables that serve as memory.
Expressing sequential composition in logic leads to long, unstructured formulas that deemphasize the specifier's intent.
The resulting specifications are difficult to maintain, and writing them is error-prone.
In addition, the specifier may need to explicitly write clauses that constrain variables to remain unchanged, in order to maintain imperative state.
This leads to longer formulas in which the intent behind individual clauses is less readable.

Another motivation relates to the temporal logic hierarchy \cite{Manna90podc,Schneider04}.
Synthesis from LTL has time complexity polynomial in the state space size, and doubly exponential in the size of the formula.
In contrast, algorithms with linear time complexity in the size of the formula are known for the fragment of generalized reactivity of rank one, known as GR(1).

In the automata hierarchy, the GR(1) fragment corresponds to an implication of deterministic B\"{u}chi automata (BAs) \cite{Piterman06,Schneider04,Morgenstern11wigp,Sohail08vmcai}.
The consequent requires some system behavior, provided that the environment satisfies the antecedent of the implication, as described in \cref{sec:games-in-logic}.
Deterministic automata can describe recurrence properties ($\always \eventually$), but not persistence ($\eventually \always$).
Intuitively, the behavior of variables uniquely determines the associated behavior of a deterministic BA.
This drops the complexity of synthesis, because the algorithm does not have to keep track of branching in the automata that is not recorded in the problem's variable.

A large subset of properties that are of practical interest in industrial applications \cite{Dwyer99icse,Manna90stanford-tr,Bloem12jcss} can be expressed in GR(1).
There do exist properties that cannot be represented by deterministic B\"{u}chi automata, e.g., persistence $\eventually \always p$.
Of these properties, those with Rabin rank equal to one are still amenable to polynomial time algorithms (by solving parity games) \cite{Ehlers11nfm}.
Higher Rabin ranks are not expected to admit polynomial time solution, unless $\textsc{P} = \textsc{NP}$ \cite{Ehlers11nfm}.
This motivates formulating the required properties in GR(1), which corresponds to Streett properties with rank one.
The winning set for a Streett objective of rank one can be computed with the same time complexity as that for a Rabin objective of rank one.
Therefore, properties in the lower Rabin ranks are known to be at least as hard to synthesize as GR(1).
This motivates formulating the required properties in GR(1), which trades off expressive power for computational efficiency.

Translating properties to deterministic automata can be done automatically, but may lead to more expensive synthesis problems than manually written properties, as reported in \cite{Morgenstern11wigp}.
So the ability to write deterministic automata directly in a structured and readable language avoids the need for automated translation, and allows fine tuning them, based on the specifier's understanding of the problem.
The trade-off is that the translation has to be performed by a human.

Another reason why specifications are not always purely declarative is that in many cases we want to synthesize a system using {\em existing} components.
In other words, we already have a {\em partial} model, which describes the possible behavior of components that already exist, e.g., because we purchased them off the self, to interface them with the part of the system that we are synthesizing.
We {\em declare} to our synthesis tool what properties the controller under design should satisfy with respect to this model.
This restricts what the system should achieve using these components, but not how exactly that will be achieved.
So the partial model is best described imperatively, whereas the goal declaratively, using temporal logic.

Educationally, the transition for students from a general purpose programming language like \textsc{Python} or \texttt{C}, directly to temporal logic constitutes a significant leap.
Using a multiparadigm language can make this transition smoother.

This work proposes a language that can describe synthesis problems for open systems that react to an adversarial environment.
The syntax is derived from that of \textsc{Promela}, whereas the semantics interprets it as a two-player turn-based game of infinite duration.
Both synchronous and asynchronously scheduled centralized systems with full information can be synthesized.
In \cref{sec:preliminaries}, we review temporal logic and relevant notions about two-player games.
The presence of two players requires declaring who controls each variable (\cref{sec:variables}), as well as the data flow, and control flow in transition systems (\cref{sec:game-graphs}).
In addition, the specification needs to be partitioned into assumptions about the environment, and guarantees that the system must satisfy (\cref{sec:preliminaries}, \cref{sec:game-graphs}).
The integration of declarative and imperative semantics is obtained by defining imperative variables (\cref{sec:variables}), deconstraining, and executability of actions (\cref{sec:guarded-statements}).
In order to be synthesized, the program is translated to temporal logic, as described in \cref{sec:translation-to-logic}.
In \cref{sec:implementation}, we discuss the implementation, and in \cref{sec:amba} significant improvements in the AMBA case study \cite{Bloem12jcss} that were possible by merging fairness requirements into a single B\"{u}chi automaton.
Relevant work is collected in \cref{sec:relevant-work}, and conclusions in \cref{sec:conclusions}.

\section{Preliminaries}
\label{sec:preliminaries}

\subsection{Linear temporal logic}

Linear temporal logic with past is an extension of Boolean logic used to reason about temporal modalities over sequences.
The temporal operators “next” $\nxt$, “previous” $\previous$, “until” $\until$, and “since” $\since$ suffice to define the other operators \cite{Pnueli77,Baier08}.
Let $AP$ be a set of variable symbols $p$ that can take values over $\booleans\triangleq\{\bot, \top\}$.
A model of an LTL formula is a sequence of variable valuations called a {\em word} $w: \naturals \rightarrow \booleans^{AP}$.
A well-formed formula is inductively defined by
$\varphi ::= p | \neg \varphi | p \wedge p | \nxt \varphi | \varphi \until \varphi | \previous \varphi | \varphi \since \varphi$.
A formula $\varphi$ is evaluated over a word $w$ at a time $i \geq 0$, and $w, i \models \varphi$ denotes that $\varphi$ holds at position $i$ of word $w$.
Formula $\nxt \varphi$ holds at position $i$ if $\varphi$ holds at position $i + 1$,
$\varphi \until \psi$ holds at $i$ if there exists a time $j \geq i$ such that $w, j \models \psi$ and for all $i \leq k < j$, it is $w, k \models \varphi$.
The operator $\eventually p \triangleq \top \until p$ requires that $p$ be “eventually” true, and the operator $\always p \triangleq \neg \eventually \neg p$ requires that $p$ be true over the whole word.
The past fragment of LTL extends it with the “previous” and “since” operators, $\previous, \since$ respectively \cite{Lichtenstein85clp,Manna89lncs,Kesten1998icalp}.
Formula $\previous \varphi$ holds at $i$ iff $i > 0$ and $w, i-1 \models \varphi$, and
formula $\varphi \since \psi$ holds at $i$ iff there exists a time $j$ with $0 \leq j \leq i$ such that $w, j \models \psi$, and for all $k$ such that $j < k \leq i$ it is $w, k \models \varphi$.
The {\em weak previous} operator $\weakprevious$ is defined as $\weakprevious \varphi \triangleq \neg \previous \neg \varphi$,
“once” $\once$ as $\once \varphi \triangleq \top \since \varphi$, and
“historically” $\historically$ as $\historically \varphi \triangleq \neg \once \neg \varphi$.
Past LTL is implemented using temporal testers \cite{Kesten1998icalp}.

\subsection{Turn-based games}

In many applications, we are interested in designing a system that does not have full control over the behavior of all variables that are used to model the situation.
Some problem variables represent the behavior of other entities, usually collectively referred to as the “environment”.
The system reads these {\em input} variables and reacts by writing to {\em output} variables that it controls, continuing indefinitely.
Such a system is called {\em open} \cite{Abadi94podc,Pnueli89popl}, to distinguish it from closed systems that have no inputs, and so full control.

The synthesis of an open system can be formulated as a two-player adversarial game of infinite duration \cite{Thomas08npgi}.
The two players in the game are usually called the protagonist (system) and antagonist (environment).
We control the protagonist, but not the antagonist.
If the players move in turns, then the game is called {\em alternating}.
Each pair of consecutive moves by the two players is called a {\em turn} of the game.
In each turn, player 0 moves first, without knowing how player 1 will choose to move in that turn of the game.
Then player 1 moves, knowing how player 0 moved in that turn.
Depending on which player we control, there are two types of game.
If the protagonist is player 1, then the game is called {\em Mealy}, otherwise {\em Moore}  \cite{Mealy55,Moore56}.
Due to the difference in knowledge about the opponent's next move between the two flavors of game, more specifications are realizable in a Mealy game, than in a Moore game.
There exist solvers for both Moore and Mealy games.
Here we will consider Mealy games only.

\subsection{Games in logic}
\label{sec:games-in-logic}

Temporal logic can be used to describe both the possible moves in a game (the {\em arena} or {\em game graph}), as well as the winning condition.
Let $\mathcal{X}$ and $\mathcal{Y}$ be two sets of propositional variables, controlled by the environment and system, respectively.
Let $\mathcal{X}'$ and $\mathcal{Y}'$ denote primed variables, where $x'$ represents the next value $\nxt x$ of variable $x$.
We abuse notation by using primed variables inside temporal formulae.

Synthesis from LTL specifications is in \textsc{2ExpTime} \cite{Pnueli89icalp,Rosner92}, motivating the search for fragments that admit more efficient synthesis algorithms.
Generalized reactivity of index one, abbreviated as GR(1), is a fragment of LTL that admits synthesis algorithms of time complexity polynomial in the size of the state space \cite{Bloem12jcss}.
GR(1) \cite{Jobstmann07cav,Pnueli10cav,Bloem10cav,Livingston12icra,Ehlers14synt} is used in the following, but the results can be adapted to larger fragments of LTL, provided that another synthesizer be used \cite{Jobstmann06fmcad,Ehlers11tacas,Ehlers2011wigp,Bohy12cav,Filippidis13list}.

The possible moves in a Mealy game can be specified by initial and transition conditions that constrain the environment and system.
Initial conditions are described by propositional formulae over $\mathcal{X} \cup \mathcal{Y}$.
Transition conditions are described by safety formulae of the form $\always \varphi_i$ where, for the environment $i=e$ and $\varphi_e$ is a formula over $\mathcal{X} \cup \mathcal{X}' \cup \mathcal{Y}$, and for the system $i=s$ and $\varphi_s$ is a formula over $\mathcal{X} \cup \mathcal{X}' \cup \mathcal{Y} \cup \mathcal{Y}'$.
Note that the system plays second in each turn, so it can see $\mathcal{X}'$, whereas the environment cannot see $\mathcal{Y}'$, because it represents future values.
The winning condition in a GR(1) game is described using progress formulae of the form $\always \eventually \psi_i, i \in \{e, s\}$, where $\psi_i$ is a propositional formula over $\mathcal{X} \cup \mathcal{Y}$.

The overall specification of a GR(1) game is of the form
\myeq{
	\Big(
		\underbrace{
			\theta_e \wedge
			\always \varphi_e \wedge
			\bigwedge_{ i = 0 }^{ n - 1 }
				\always \eventually \psi_{e, i}
		}_{assumption}
	\Big)
		\strictarrow
	\Big(
		\underbrace{
			\theta_s \wedge
			\always \varphi_s \wedge
			\bigwedge_{ j = 0 }^{ m - 1 }
				\always \eventually \psi_{s, j}
		}_{assertion}
	\Big)
}
Note that requirements that constraint the environment are called {\em assumptions} and guarantees that the system must satisfy are called {\em assertions}.
Assumptions limit the set of admissible environments, because, in practice, it is impossible to satisfy the design requirements in arbitrarily adversarial environments \cite{Abadi94podc}.
The {\em strict realizability} implication $\strictarrow$ above is interpreted by prioritizing between safety and liveness \cite{Bloem12jcss}, to prevent the system from violating the safety assertion, in case this would allow it to prevent the environment from satisfying the liveness assumption.
The GR(1) synthesis algorithm has time complexity $O\big(n m \abs{\Sigma}^2\big)$ \cite{Bloem12jcss},
where $n$ ($m$) is the number of recurrence assumptions (assertions),
and $\Sigma$ the set of all possible variable valuations.

\section{Language definition}

The language we are about to define is syntactically an extension of \textsc{Promela} \cite{Holzmann03}, but its semantics is defined by a translation to turn-based infinite games with full information.
\textsc{Promela} is a guarded command language that can represent transition systems, non-deterministic execution, and guard conditions for determining whether statements are executable \cite{Holzmann03,Dijkstra75cacm}.
Its syntax can be found in the language reference manual \cite{Holzmann03,spinroot}.
Here we will introduce syntactic elements only as needed for the presentation.
Briefly, we mention that a program comprises of transition systems and automata, whose control flow can be described with sequential composition, selection and iteration statements, \texttt{goto}, as well as blocks that group statements for atomic execution.

\subsection{Variables}
\label{sec:variables}

\paragraph{Ownership}

In a game, variables from $\mathcal{X}$ are controlled by the environment and variables from $\mathcal{Y}$ by the system.
We call {\em owner} of a variable the player that controls it.
We use the keywords \texttt{env} and \texttt{sys} to signify the owner of a variable.
Variables can be of Boolean, bit, byte, (bounded) integer, or bitfield type.

\paragraph{Declarative and imperative semantics}

In imperative languages, variables remain unchanged, unless explicitly assigned new values.
In declarative languages, variables are free to change, unless explicitly constrained \cite{vanRoy04}.
In verification, both declarative languages like \textsc{TLA} \cite{Lamport02} and \textsc{SMV} \cite{Cavada10nusmv} have been used, as well as imperative languages like \textsc{Promela} and \textsc{Dve} \cite{Barnat13cav}.

In a synthesis problem, there are variables that are more succinct to describe declaratively, whereas others imperatively.
For this reason, we combine the two paradigms, by introducing a new keyword \texttt{free} to distinguish between imperative and declarative variables.
Variables whose declaration includes the keyword \texttt{free} are by default allowed to be assigned any value in their domain, unless explicitly constrained otherwise.
Variables without the keyword \texttt{free} have imperative semantics, so their value remains unchanged, unless otherwise explicitly stated.
Let $V^{free}$ denote free variables, and $V^{imp}$ imperative variables, and $V_p$ the variables of player $p \in \{e, s\}$.

\paragraph{Ranged integer data type}

Symbolic methods for synthesis use reduced ordered binary decision diagrams (BDDs) \cite{Bryant86tc,Baier08}, which represent sets of states, and relations over states.
As operations are performed between BDDs, these can grow quickly, consuming more memory.
The growth can be ameliorated by reordering the variables over which a BDD is defined.
Reordering variables can be prohibitively expensive, as discussed in \cref{sec:measurements}, so reducing the number of bits is a primary objective.
In addition, the complexity of GR(1) synthesis is polynomial in the number $\abs{\Sigma}$ of variable valuations, which grows exponentially with each additional bit.
We can reduce the number of bits by using bitfields whose width is tailored to the problem at hand.
For convenience, the {\em ranged} integer type \texttt{int(MIN, MAX)} is introduced to define a variable $x \in \left\{\mathrm{MIN}, \mathrm{MIN} + 1, \dots, \mathrm{MAX}\right\}$, with saturating semantics \cite{Gennari07cmu}.
An integer with saturating semantics cannot be incremented when its value reached the maximal in its range, i.e., $\mathrm{MAX}$.

A ranged integer is represented by a bitfield.
The bitfield is comprised of bits, so it can only range between powers of two.
The ranged integer though may have an arbitrary range.
For this reason, safety constraints are automatically imposed on the bitfield representing the ranged integer.
In other words, if $x$ is the integer value of the bitfield, and it represents an integer that can take values from $\mathrm{MIN}$ to $\mathrm{MAX}$, then the constraint $\always (\mathrm{MIN} \leq x' \leq \mathrm{MAX})$ is added to the safety formula, and $\mathrm{MIN} \leq x \leq \mathrm{MAX}$ to the initial condition of the player that owns the ranged integer.

Other numerical data types have mod wrap semantics.
The value of an integer with mod wrap semantics overflows to $\mathrm{MIN}$ (underflows to $\mathrm{MAX}$) if incremented when equal to the maximal value $\mathrm{MAX}$ (minimal value $\mathrm{MIN}$).
Mod wrap semantics are available only for integers that range over all values of a (signed) bitfield, because the modulo operation would otherwise be needed.
Any BDD describing a modulo operation is at best of exponential size \cite{Bryant86tc}.

\subsection{Programs representing games}
\label{sec:game-graphs}

\begin{figure}
	\begin{subfigure}{0.24\textwidth}
		\centering
		\includesvg[width=\textwidth]{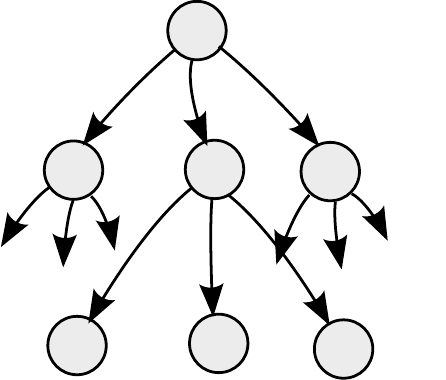}
		\caption{\texttt{assume env}}
	\end{subfigure}
	\begin{subfigure}{0.24\textwidth}
		\centering
		\includesvg{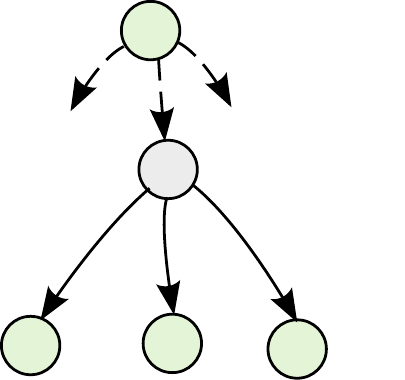}
		\caption{\texttt{assume sys}}
	\end{subfigure}
	\begin{subfigure}{0.24\textwidth}
		\centering
		\includesvg{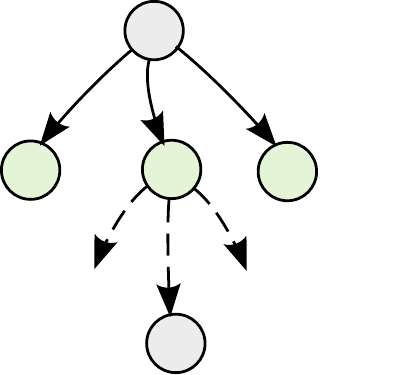}
		\caption{\texttt{assert env}}
	\end{subfigure}
	\begin{subfigure}{0.24\textwidth}
		\centering
		\includesvg{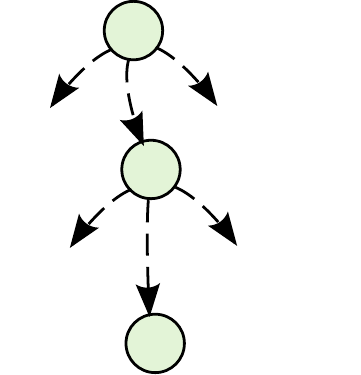}
		\caption{\texttt{assert sys}}
	\end{subfigure}
	\caption{An assumption (assertion) process constrains the environment (system) variables, and \texttt{env} (\texttt{sys}) declares who chooses the next statement to be executed (when there are multiple).}
	\label{fig:flow-quantification}
\end{figure}

In many synthesis problems, the specification includes graph-like constraints.
These may originate from physical configurations in robotics problems, deterministic automata to express a formula in GR(1), or describe abstractions of existing components that are to be controlled.
These constraints can be described by processes.
A process describes both control and data flow.
In order to discuss control and data flow, we will refer to {\em program graphs}.
A program graph is an intermediate representation of a process, after parsing and control flow analysis.
For our purposes, a {\em program graph} is a rooted directed multi-graph $P_r \triangleq (V_r, E_r)$ whose edges $E_r$ are labeled with program statements, and nodes $V_r$ abstract states of the system \cite{Keller76cacm,Baier08}.
Execution starts from the graph's root.
A multi-digraph is needed, because, between two given nodes, there may exist edges labeled with different program statements.

Control flow is the traversal of edges in a program graph (i.e., execution of statements), whereas data flow is the behavior of program variables along this traversal.
A {\em program counter} $\mathit{pc}_r$ is a variable used to store the current node in $V_r$.
A natural question to ask is who controls the program counter.
Another question is whose data flow is constrained by the program graph $P_r$.
In the next section, we define syntax that allows declaring the player that controls the program counter, and the player that is constrained to manipulate the variables it owns, according to the statements selected by the program counter.
This allows defining both processes where control and data flow are controlled and constrain the same player, but also processes with mixed control.
If one player controls the program counter, and the opponent reacts by choosing a compliant data flow, then the process itself describes a game.

As an example, suppose that for a given process, the environment controls the program counter, and the system the local program variables.
By choosing the next value of the program counter, the environment selects the next program statement that will execute.
The system must react by choosing the next values of the local variables, such that they satisfy the selected program statement.
The environment can select as next program statement any statement that is satisfiable by a system reaction, as discussed in more detail in \cref{sec:guarded-statements}.
But other constraints, e.g., LTL formulae, can prevent the system from satisfying this statement.

The consecutive assignments of values to variables by the environment and the system can be represented by a {\em game graph}.
Each node in the game graph corresponds to a valuation of variables.
From each node, a single player can assign new values to its variables, depending on the outgoing edges at that node.
The choice of outgoing edge at environment (system) nodes is universal (existential).
The nodes in a game graph correspond to universal and existential nodes in alternating tree automata \cite{Chandra81jacm,Muller86alp,Vardi95cst,Vardi96lc,Kupferman05focs}.
Nondeterminism with universal quantification is known as {\em demonic}, \cite[p.85]{Hoare85csp}, \cite{Sondergaard1992cj}, otherwise as {\em angelic} \cite{McCarthy63,Floyd67jacm,Broy86tcs}.
In the previous example, the environment's choice of control flow occurs at universal nodes in the game graph.
The system's reaction, by assigning to local variables, occurs at existential nodes in the game graph.
The correspondence of control and data flow with a game graph is shown in \cref{fig:flow-quantification}.

\subsubsection{Syntax}

Program graphs are declared with the \texttt{proctype} keyword of \texttt{Promela} followed by statements enclosed in braces.
The keyword \texttt{assume} (\texttt{assert}) declares a process that constrains the environment's (system's) data flow.
These keywords are common in theorem proving and program verification languages \cite{Leino10lpair}.

\begin{wrapfigure}{r}{6cm}
\centering
\includesvgpdf[width=5cm]{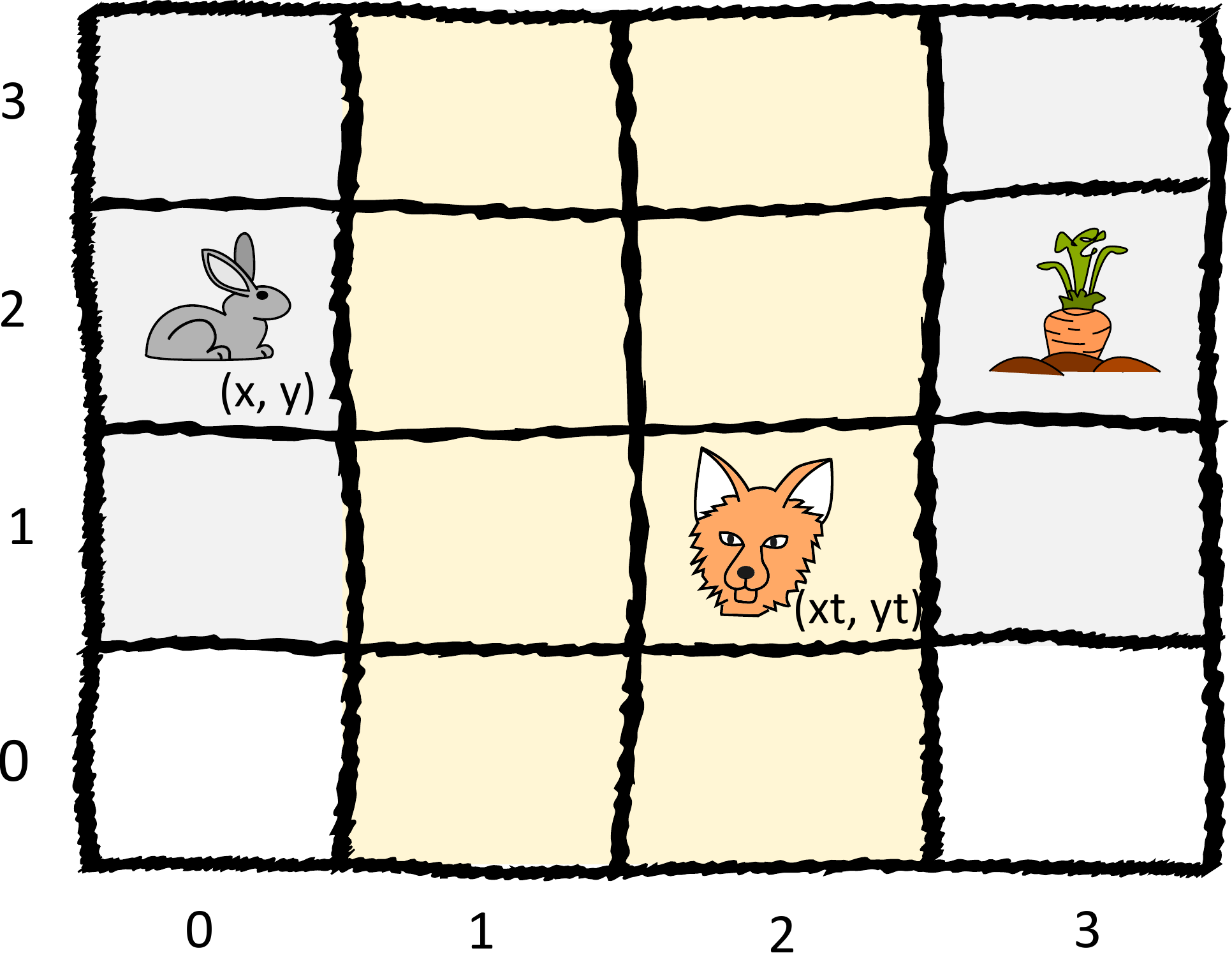}
\caption{Adversarial game.}
\label{fig:board_game}
\end{wrapfigure}

The keyword \texttt{env} (\texttt{sys}) declares that the environment (system) controls the program counter $\mathit{pc}_r$ of a process, \cref{fig:flow-quantification}.
The implementation of \texttt{assume sys} is the most interesting, and is described in \cref{sec:translation-to-logic}.
We will call program graphs {\em processes}, noting that these processes have full information about each other, so they correspond to centralized synthesis, not distributed.
The program counter {\em owner} is the player that controls variable $\mathit{pc}_r$.
The {\em process player} is the player constrained by the program graph.

\paragraph{Example}

For example, the specification in \cref{spec:board_game} defines a game between two players: the Bunny, and the Fox, that move in turns, as depicted in \cref{fig:board_game}.
Each logic time step includes a move by the Fox from $(x_t, y_t)$ to $(x_t' , y_t')$, followed by a move by the Bunny from $(x, y)$ to $(x', y')$.
The Bunny must reach the carrot, without moving through a cell that Fox is in (\texttt{assert ltl}).
The Fox can only move between $x_t \in \{1, 2\}$, and has to keep visiting the lower row (\texttt{assume ltl}).
The Fox can move diagonally, but the Bunny only vertically or horizontally.
Both players have an option to stay still (\texttt{skip}).
Note that $x_t$ is a declarative variable, so it can change unless constrained.

\begin{lstlisting}[basicstyle=\ttfamily\small,xleftmargin=3em,multicols=2,caption={Simple example.},label={spec:board_game}]
#define H 3

free env int(1, 2) xt;
env int(0, H) yt;

assume env proctype fox(){
    do
    :: yt = yt - 1
    :: yt = yt + 1
    :: skip
    od
}

assume ltl { []<>(yt == 0) }

sys int(0, 3) x;
sys int(0, H) y;

assert sys proctype bunny(){
    do
    :: x = x - 1
    :: x = x + 1
    :: y = y - 1
    :: y = y + 1
    :: skip
    od
}

assert ltl {
  [] ! ((x == xt) && (y == yt)) &&
  [] ! ((xt' == x) && (yt' == y)) &&
  []<>((x == 3) && (y == 2)) }
\end{lstlisting}

The process \texttt{fox} constrains the environment variables $x_t, y_t$ (\texttt{assume}) and the environment controls its program counter (\texttt{env}).
The \texttt{do} loops define alternatives that each player must choose from to continue playing the game.
Note that nondeterminism in process \texttt{fox} is demonic (universally quantified), whereas in \texttt{bunny} angelic (existentially quantified), i.e., the design freedom given to the synthesis tool.
Each player has full information about all variables in the game, both local, as well as global, and auxiliary.
The solution is a strategy represented as a Mealy transducer \cite{Mealy55} that the Bunny can use to win the game.

The conjuct
$\varphi \triangleq \always \neg ((x = x_t') \wedge (y = y_t'))$
prevents the Bunny from moving next to the Fox, from where the Fox can catch it in the next turn.
The $\always$ operator requires that, at each time step, the formula $\neg$ as main operator be true.

\subsection{Statements}
\label{sec:guarded-statements}

Control flow can be defined using selection (\texttt{if}) and repetition (\texttt{do}) statements, \texttt{else}, \texttt{break}, \texttt{goto}, and labeled statements.
The statements \texttt{run}, \texttt{call}, \texttt{return} are not supported, because dynamic process creation would dynamically add BDD variables.
In this section, we define expressions, assignments, and their executability.

\paragraph{Expressions}
\label{sec:expressions}

Primed variables (that correspond to using the “next” operator $\nxt$)
can appear in expressions to refer to the “next” values of those variables, as in the syntax of synthesis tools and \textsc{TLA} \cite{Lamport94tpls}.
The operators weak previous $\weakprevious$ and strong previous $\previous$ are expressed with the tokens \texttt{-X} and \texttt{--X}, respectively.
Following \textsc{TLA}, we will call {\em (state) predicate} an expression that contains no primed variables and {\em action} an expression that contains primed variables \cite{Lamport94tpls}.
Actions can be regarded as generalized assignments, in a sense that will be made precise later.
Primed system variables cannot appear in assumption processes, because they refer to values not yet known to the environment.
Using a GR(1) synthesizer as back-end, multiple priming within a single statement is not allowed, but can be allowed if a full LTL synthesizer is used as back-end \cite{Jobstmann06fmcad,Finkbeiner13sttt}.

\paragraph{Deconstraining}

By default, imperative variables are constrained to remain invariant.
If any assumption (assertion) process executes a statement that contains a primed environment (system) variable, then that variable is not constrained to remain unchanged in that time step.
For example, in the assertion \texttt{sys bit x = 0; (x == 0); (x' == 1 - y)} the variable \texttt{x} is constrained by $x' = x$ when \texttt{x == 0} is executed, but the synthesizer is allowed to pick its next value as needed, in order to satisfy \texttt{x' == 1 - y}.
Note that statements in assumption (assertion) processes that contain primed imperative system (environment) variables do {\em not} deconstrain those variables, because assumptions (assertions) are relevant only to the environment's (system's) data flow.

\paragraph{Assignments}

In \textsc{Promela}, expressions are evaluated by first converting all values to integers, then evaluating the expression with precision that depends on the operating system and processor, and updating the assigned variable's value, truncating if needed.
Let $\mathrm{trunc}(y, w)$ denote a function that truncates the value of expression $y$ to bitwidth $w$.
An assignment \verb+x = expr+ is translated to the logic formula $x' = \mathrm{trunc}(\mathit{expr}, \mathrm{width}(x))$, if variable $x$ has mod wrap semantics, and to $x' = \mathit{expr}$ otherwise.
If variable $x$ is imperative, then it is deconstrained.

\paragraph{Statement executability}

A condition called {\em guard} is associated to each statement \cite{Dijkstra75cacm}.
The process can execute a statement only if the guard evaluates to true.
If a process currently has no executable statement, then it {\em blocks}.
For each statement, its guard is defined by existential quantification of the primed variables of the data flow player.
The quantification is applied after the statement is translated to a logic formula.
So the guard of a statement is the realizability condition for that statement.
It means that, from the local viewpoint of that statement only, given the current values of variables in the game, the constrained player can choose a next move.
So the scheduler cannot pick as next process to execute a process that has blocked.
Clearly, if all processes block, then that player deadlocks.

Using this definition, the guard of a state predicate is itself, as in \textsc{Promela}.
The implementation quantifies variables using the \textsc{Python} binary decision diagram \texttt{dd} \cite{Filippidis15github-dd}.
If an unsatisfiable guard is found, then the implementation raises a warning.
For example, if we inserted the statement \verb+xt && xt' && y'+ in the process \texttt{bunny} (see example), then its guard would be $\exists y'. x_t \wedge x_t' \wedge y' = x_t \wedge x_t'$.
Similarly, the guard of an expression \verb+xt && xt'+ in the process \texttt{fox} is $x_t$.

\def\ps{\mathit{ps}}
\def\pc{\mathit{pc}}
\def\key{\mathrm{key}}
\def\ex{\mathit{ex}}
\def\gids{\mathrm{gids}}
\def\pids{\mathrm{pids}}
\def\oute{\mathrm{oute}}
\def\guard{\mathrm{guard}}
\def\inv{\mathrm{inv}}
\def\pmt{\mathit{pm}}

\section{Translation to logic}
\label{sec:translation-to-logic}

In this section, we describe how a program is translated to temporal logic, in particular GR(1).
For each process, the starting point is its program graph, which has edges labeled by program statements, and describes the control flow of a process in the source code.
The construction of program graphs from source code is the same as for \textsc{Promela} \cite{Holzmann03}, and described in detail in \cite{Filippidis15cds3-synt}.

\begin{figure}
\begin{subfigure}{0.6\textwidth}
\centering
\includesvgpdf[width=\textwidth]{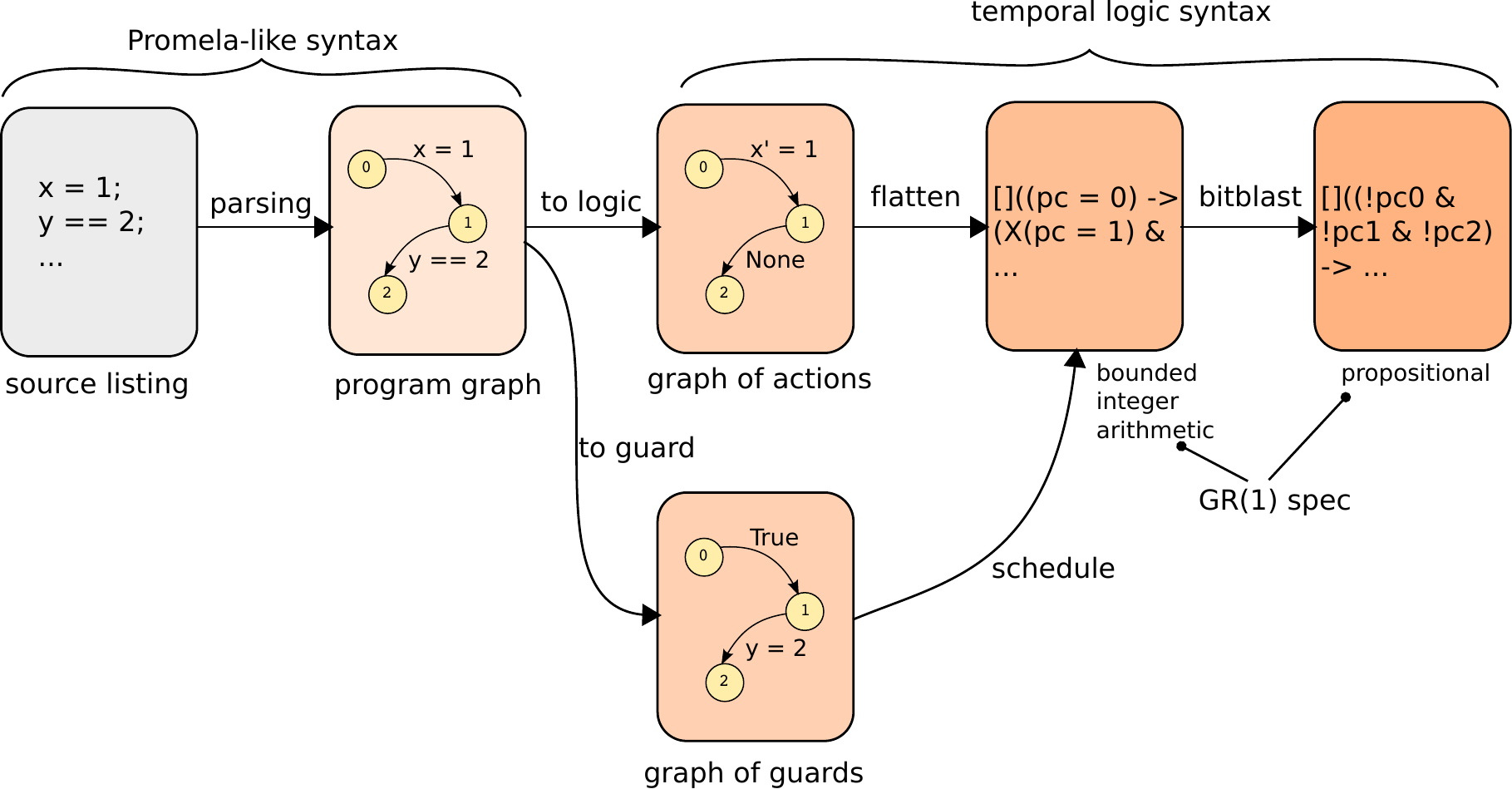}
\caption{Compiler architecture.}
\label{fig:compiler_architecture}
\end{subfigure}
\begin{subfigure}{0.35\textwidth}
\centering
\includegraphics[width=\textwidth]{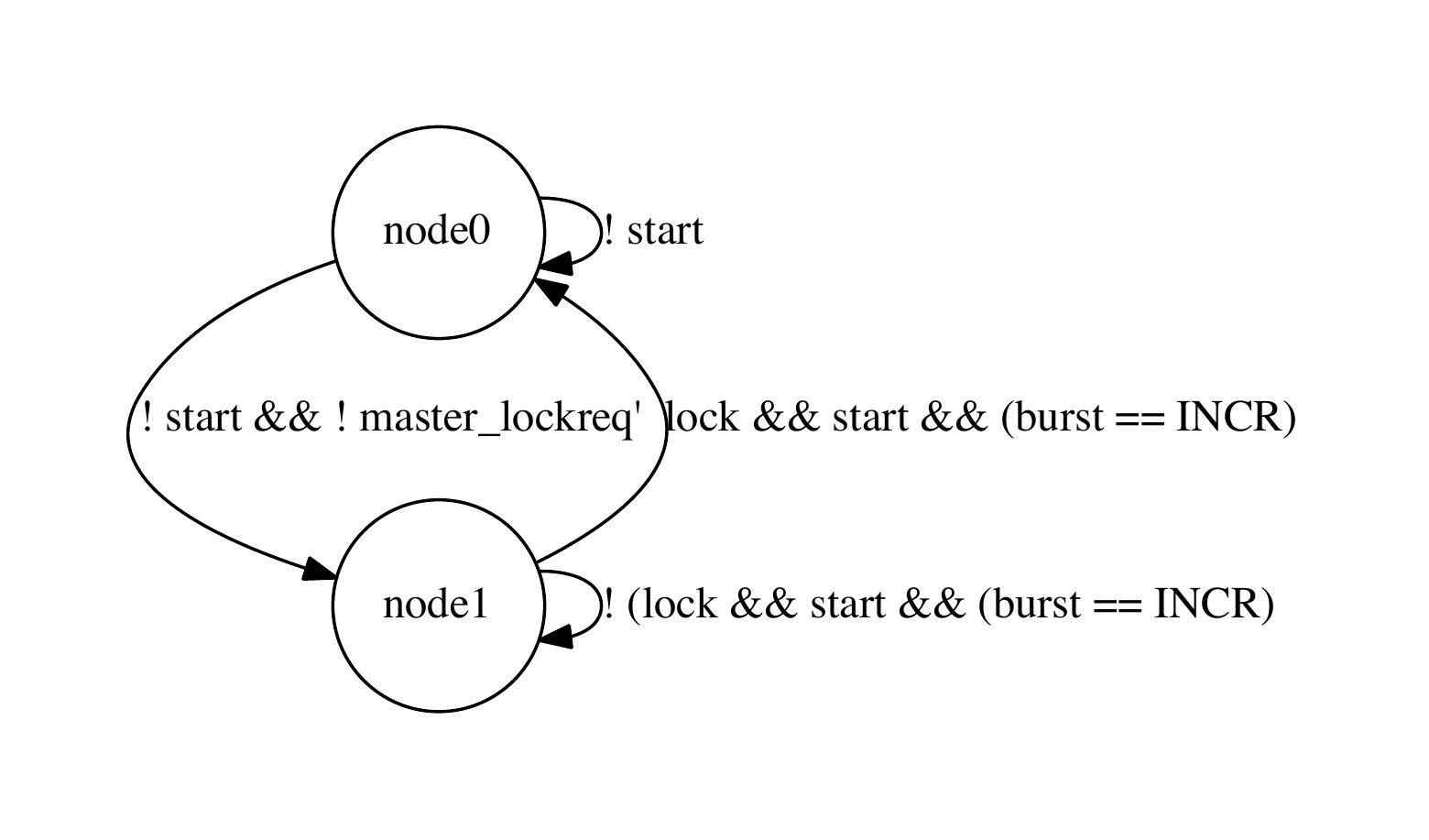}
\caption{Program graph that corresponds to process \texttt{maintain\_lock} in \cref{code:amba}.}
\label{fig:aut_g2}
\end{subfigure}
\caption{Compiling programs to temporal logic.}
\end{figure}

Here we give a brief example.
Consider the process \texttt{maintain\_lock} in \cref{code:amba}.
It has two \texttt{do} loops, with two outgoing edges each.
The corresponding program graph is shown in \cref{fig:aut_g2}.
Each statement labels one edge, and that edge can be traversed if the guard associated to the statement evaluates to true.
The guard can contain primed variables, requiring that the dataflow player manipulates them so as to make the edge's guard true.
Otherwise, the player cannot traverse an edge with false guard.
This program graph is translated further to logic, as described next.
The semantics of the language are defined by this translation to logic.

There are three groups of elements in a program: processes, \texttt{ltl} blocks, and the scheduler that picks processes for execution.
The scheduler is not present in the source code, but is added during translation, to represent the products between processes.
The translation can be organized into a few thematically related sets of formulae.
Due to lack of space, we are going to discuss the most interesting and representative of these at a high level.
The full translation can be found in \cite{Filippidis15cds3-synt}, and in the implementation.
There are four groups of formulae:
(i) control and data flow,
(ii) invariance of variables,
(iii) process scheduler,
(iv) exclusive execution (atomic).

\paragraph{Control and data flow}

The translation of processes is reminiscent of symbolic model checking \cite{McMillan92phd}, but differs in that there are two players, and both play in each logic time step.
This requires carefully separating the formulae into assumptions and guarantees (assertions).

Suppose that the scheduler selects process $r$ to execute (how is explained later).
At a given time step, a process is at some node $i$ in its program graph, and will transition to a next node $j$, by traversing an edge labeled by a program statement.
The player that controls the program counter $\pc_r$ selects the next statement, so the edge in the program graph.
The player that is constrained by that process has to make sure that it {\em complies}, by picking values for variables that it controls such that the statement is satisfied.
Recall that the scheduler can only pick from processes that have a satisfiable statement, so whenever a process executes, there will exist a satisfiable next statement.
Of course, conflicts can arise between different synchronous processes that can lead to deadlock, and it is the synthesizer's task to avoid such situations, to avoid losing the game.
The transition constraints are encoded by the formula
\myeq{
\label{eq:trans}
&\mathrm{trans}(p, r)
\triangleq \bigwedge_{i \in N_r}
\big(
	(\pc_r = i)
		\rightarrow
	\bigvee_{ (i, j, k) \in E_r}
	\begin{array}{c}
		\varphi_{r, i, j, k} \wedge
		(\tilde{\pc}_r = j) \wedge
		(\tilde{\key}_r = k) \wedge
		\mathrm{exclusive}(p, r, i, j, k)
	\end{array}
\big)
}
where $N_r$ denotes the set of nodes, and $E_r$ the multi-edges of the program graph of a process, $p$ denotes the player ($e, s$).
The logic formula equivalent to bitblasting the statement labeling edge $(r, i, j, k)$ is $\varphi_{r, i, j, k}$.
For \texttt{assume sys} processes, the system selects the next edge one time step before the scheduler decides whether that environment process will execute, so two copies are needed, $\pc_r, \hat{\pc}_r$ (system variables).
So in an \texttt{assume sys} process,
$\tilde{\pc}_r \triangleq \hat{\pc}_r,
\tilde{\key}_r \triangleq \key(r)$, and in other processes
$\tilde{\pc}_r \triangleq \pc_r',
\tilde{\key}_r \triangleq \key(r)'$.
The variable $\key(r)$ selects among multi-edges, and is controlled by the same player as the program counter $\pc_r$ of the process with pid $r$.
In a system process, if node $j$ is in an atomic block, then $\mathrm{exclusive}(p, r, i, j, k)$ sets the auxiliary variables $\ex_s'$ and $\pmt_s'$ to request atomic execution from the scheduler.
The integer variable $\ex_s$ stores the identity of the process that requests atomic execution, and the bit $\pmt_s$ requests that the environment be preempted, if the scheduler grants the request for atomic execution.
\myeq{
\mathrm{dataflow}(r)
&\triangleq (\ps(r)' = m(r)) \rightarrow \mathrm{trans}(p, r)
\\
\mathrm{selectable}(r)
&\triangleq \mathrm{blocked}(r) \rightarrow (\ps(r)' \neq m(r))
\\
\mathrm{control\_flow}(r)
&\triangleq \mathrm{ite}\big(
(\ps(r)' = m(r)),\;
\mathrm{pc\_trans}(p, r),\;
\inv(\pc_r)
\big)
\\
\mathrm{blocked}(r)
&\triangleq \bigvee_{i \in N_r}
\big(
(\pc_r = i) \wedge
\bigwedge_{ (i, j, k) \in E_r }
	\neg \guard_{r, i, j, k}
\big)
}
The environment variables $\ps(r)$ select the process or synchronous product that will execute next inside an asynchronous product (top context is an asynchronous product).
For this purpose, each process and product have a local integer id $m(\cdot)$ among the elements inside the product that contains them.
The transition relation for the program counter depends on the type of process.
For \texttt{assume sys} processes,
$\mathrm{pc\_trans}(p, r) \triangleq (\pc_r' = \widehat{\pc}_r)$, and for other processes it equals $\mathrm{guards}(r)$.
The condition $\mathrm{guards}(r)$ constrains the program counter to follow unblocked edges in a process.
It is necessary when the control and data flow are controlled by different players, because whoever moves the program counter, can otherwise pick an edge with a statement that blocks the other player.
In addition, for \texttt{assume sys} processes, a separate constraint with same form as $\mathrm{guards}(r)$, but different priming of sub-expressions is imposed on the program counter copy $\widehat{\pc}_r$.
The ternary conditional is denoted by $\mathrm{ite}(a, b, c)$.

\paragraph{Invariance of variables}

When a process is not executing, its declarative local variables must be constrained to remain invariant ($x' = x$).
Also, imperative variables must remain invariant whenever no process executes a statement (edge) that either is an expression and contains a primed copy of that variable, or is an assignment.
These are ensured by the following equations
\myeq{
&\mathrm{local\_free}(p, r)
\triangleq
	(\ps(r)' \neq m(r))
		\rightarrow
	\bigwedge_{x \in V^{\text{free}}_{p, r} } \inv(x)
\\
&\mathrm{imperative\_inv}(p, r)
\triangleq
\mathrm{array\_inv}(p, r) \wedge
\bigwedge_{x \in V^{\text{imp}}_{p, r} } (
	\inv(x) \vee
	\bigvee_{ (i, j, k) \in E_r. x \in \mathrm{deconstrained}(r, i, j, k) }
		\mathrm{edge}(r, i, j, k)
)
}
where $V^{\text{free}}_{p, r}$ are free local variables of player $p$ in process $r$.
For \texttt{assume sys} processes, it is
$\mathrm{edge}(r, i, j, k)
\triangleq
(\ps(r)' = m(r)) \wedge
(\pc_r = i) \wedge
(\hat{\pc}_r = j) \wedge
(\key(r) = k)
$, and for other types of processes
$\mathrm{edge}(r, i, j, k)
\triangleq
(\ps(r)' = m(r)) \wedge
(\pc_r = i) \wedge
(\pc_r' = j) \wedge
(\key(r)' = k)
$.
Primed references to elements in imperative arrays deconstrain only the referenced array element, ensured by $\mathrm{array\_inv}$.

\paragraph{Scheduler}

The scheduler (environment) has to select the processes that will execute.
Products of processes can be defined in the source code by enclosing processes, or other products, in braces preceded by the keywords \texttt{async} and \texttt{sync}.
They can be nested.
\texttt{async} defines an asynchronous product, and the scheduler picks some unblocked process or product inside it to execute next.
If all processes/products in an asynchronous product $k$ have blocked, then the scheduler sets the corresponding variable $\ps_k$ to a reserved value ($n_k$).
The reserved value is also used if the asynchronous product is nested in a synchronous product that currently is not selected to execute.

If the top product blocks, then the player has deadlocked, losing the game.
The only exception is when the environment is preempted by a request from a system process for atomic execution.
At a high level, this behavior is expressed as follows for scheduling the environment processes.
\myeq{
\mathrm{product\_selected}(k)
&\triangleq
(\ps(k)' \neq m(k))
\leftrightarrow
( \ps_k' = n_k )
\\
\mathrm{selectable\_element}(r)
&\triangleq
\mathrm{element\_blocked(r)}
\rightarrow
(\ps(r)' \neq m(r))
\\
\mathrm{element\_blocked}(r)
&\triangleq
\begin{cases}
\mathrm{blocked}(r),\; \text{if $r$ is a process}
\\
\mathrm{sync\_blocked}(r),\; \text{if $r$ is a synchronous product}
\\
\mathrm{async\_blocked}(r),\; \text{if $r$ is an asynchronous product}.
\end{cases}
}
\myeq{
\mathrm{sync\_blocked}(k)
&\triangleq
\bigvee_{ r \in R_k }
\mathrm{element\_blocked}(r)
\\
\mathrm{async\_blocked}(k)
&\triangleq
\bigwedge_{ r \in R_k }
\mathrm{element\_blocked}(r)
}
\myeq{
\mathrm{pause\_env\_if\_req}
&\triangleq
(\ps_{\mathrm{env\_top}}' = n_e)
\leftrightarrow
(\pmt_s \wedge (\ps_{\mathrm{sys\_top}}' = \ex_s < n_s)).
}
The expression $\mathrm{element\_blocked}(r)$ depends on the $\mathrm{blocked}(z)$ expressions, and ensures that the scheduler doesn't select a synchronous product containing some blocked process, neither an asynchronous product where all processes are blocked.
Recall also $\mathrm{selectable}$ from earlier, which applies to individual processes.
Analogous formulae apply to system processes.
For system processes, the top-level asynchronous product implication in $\mathrm{async\_blocked}(r)$ must be replaced with equivalence, to force the environment to choose some system process (or product) to execute, when there exist unblocked ones.
Note that the asynchronous products here are in the context of full information, so the system is {\em not} asynchronous in the sense of \cite{Klein12vmcai,Pnueli09memocode}.

\paragraph{Exclusive execution}

A system process of the top asynchronous product can request to execute atomically by setting the variables $\pmt_s, \ex_s$, \cref{eq:trans}.
If that process remains unblocked in the next time step, then the scheduler will grant it uninterrupted execution, until it exits atomic context (either blocked, or reached statements outside the \texttt{atomic\{...\}} block).
\myeq{
&\mathrm{grant}_s
\triangleq \bigwedge_{ r \in \pids(s) }\big(
	((\ex_s = m(r)) \wedge \mathrm{frozen\_unblocked}(r) )
		\rightarrow
	(\ps(r)' = m(r))
\big)
}
Recall also that the environment is allowed to pause only if preempted by the system, otherwise it loses the game ($\mathrm{pause\_env\_if\_req}$).
The formula $\mathrm{frozen\_unblocked}(r)$ checks whether the system would block, in case the environment froze, granting it exclusive execution.
In case the system will block, then the request is not granted, and atomicity lost.
This requires substituting primed environment variables with unprimed ones, as follows
\myeq{
\mathrm{frozen\_unblocked}(r)
&\triangleq
\begin{cases}
\bigvee\limits_{ i \in N_r }
\big(
(\pc_r = i) \wedge
\bigvee\limits_{ (i, j, k) \in E_r }
\mathrm{guard\_test}(r, i, j, k)
\big)
,\; \text{if $\mathrm{player}(r) = s$}
\\
\neg \mathrm{blocked}(r),\; \text{otherwise}
\end{cases}
\\
\mathrm{guard\_test}(r, i, j, k)
&\triangleq
\begin{cases}
\mathrm{guard}(r, i, j, k)\vert_{\text{ $x / x'$ for $x \in \mathcal{X}$ }},
\; \text{if $i$ in atomic context}
\\
\mathrm{guard}(r, i, j, k),\; \text{otherwise}.
\end{cases}
}
The reason is that this formula corresponds to the case that the environment sets $x' = x$ for the program variables it owns.
If atomicity is lost in this turn, then the environment does not need to set $x' = x$, and this is ensured by the definition of $\mathrm{guard\_test}(r, i, j, k)$.

As in \textsc{Promela}, LTL formulae that express safety are deactivated during atomic execution (in implementation, an option allows making atomic execution visible to LTL properties).
They are re-activated as soon as atomicity is lost.
\myeq{
&\mathrm{mask\_env\_ltl}
\triangleq
\mathrm{ite}(
\pmt_s \wedge (\ps_{\mathrm{sys\_top}}' = \ex_s < n_s),\;
\mathrm{freeze\_env\_free},\;
\psi_{\text{env safety \texttt{ltl}}}
)
}
For the system, $\mathrm{mask\_sys\_ltl}$ is defined similarly.
The formula $\mathrm{freeze\_env\_free}$ constrains declarative environment variables in global context and inside system processes to remain unchanged while the system is granted exclusive execution.

If an \texttt{atomic} block appears in a process, then the \texttt{ltl} properties in the program must not contain primed variables, to ensure that the above translation yields the intended interpretation (stutter invariance).
If unbounded loops appear inside an atomic context, then there can be no liveness assumptions.
The reason is that the system can “hide” in atomic execution forever, preventing the environment from satisfying its liveness assumptions, thus winning trivially.
In order to avoid this, the environment liveness goals must be disjoined with strong fairness, a persistence property ($\eventually \always$), which is outside of the GR(1) fragment.
An extension to use a full LTL synthesizer is possible, though not expected to scale as well.
Labels in the code that contain “progress” result in accepting states (liveness conditions).
Those expressions described but not defined above, the initial conditions, and a listing into assumptions and assertions can be found in the technical report \cite{Filippidis15cds3-synt}.

\section{Implementation}
\label{sec:implementation}

The implementation is written in \textsc{Python} and available \cite{Filippidis15github-openpromela,Filippidis15github-promela,Filippidis15github-omega} under a BSD license.
The frontend comprises of a parser generator that uses \textsc{Ply} (Python \texttt{lex}-\texttt{yacc}) \cite{ply34doc}.
The parser for the proposed language subclasses and extends a separate parser for \textsc{Promela} \cite{Filippidis15github-promela}, to enable use of the latter also by those interested in verification activities.
After parsing and program graph construction, the translation described in \cref{sec:translation-to-logic} is applied \cite{Filippidis15github-openpromela}.
This results in linear temporal logic formulae that contain modular integer arithmetic.
At this point, each \texttt{ltl} block is syntactically checked to be in the GR(1) fragment, and split into initial condition, action, and recurrence conjuncts \cite{Filippidis15github-omega}.
The past fragment is then translated using temporal testers \cite{Kesten1998icalp}.
In the future, the syntactic check can be removed, and a full LTL synthesis algorithm used.

The next step encodes signed arithmetic in bitvector logic using two's complement representation \cite{Kroening08}.
The resulting formulae are in the input syntax recognized by the \textsc{Slugs} synthesizer \cite{Ehlers14synt}.
This prefix syntax includes {\em memory buffers}, which enable avoiding repetition of formulae.
For example, \verb+$ 3 x a b  & ?1 ?0  | ?2 ! ?0 + describes the ternary conditional $\mathrm{ite}(x, a, b)$.
Memory buffers prevent the bitblasted formulae from blowing up.
The \textsc{Slugs} distribution includes an encoder of unsigned addition and comparison into bitvector logic using memory buffers.
Here, signed arithmetic and arrays are supported.
The bitblaster code is a separate module, which can be reused as a backend to other frontends.
The resulting formula is passed to the \textsc{Slugs} synthesis tool to check for realizability and construct a winning strategy as a Mealy transducer.

\section{AMBA AHB Case study}
\label{sec:amba}

\begin{figure}
\centering
\begin{subfigure}{0.8\textwidth}
\includegraphics[width=\textwidth]{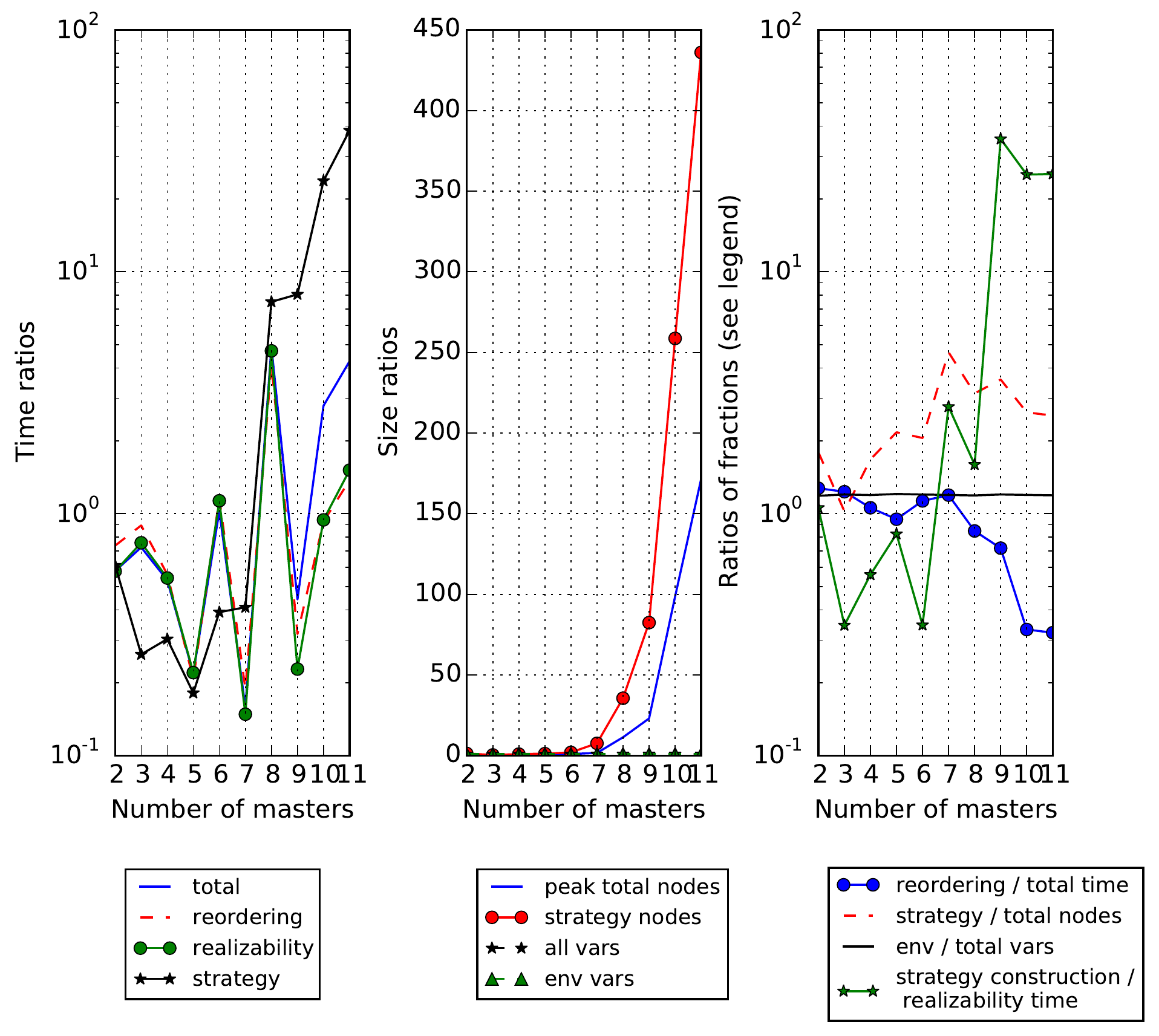}
\caption{Conjunction divided by BA, no reordering.}
\label{subfig:comparison_ba_mine_no_conj_mine_no}
\end{subfigure}
\begin{subfigure}{0.6\textwidth}
\includegraphics[width=\textwidth]{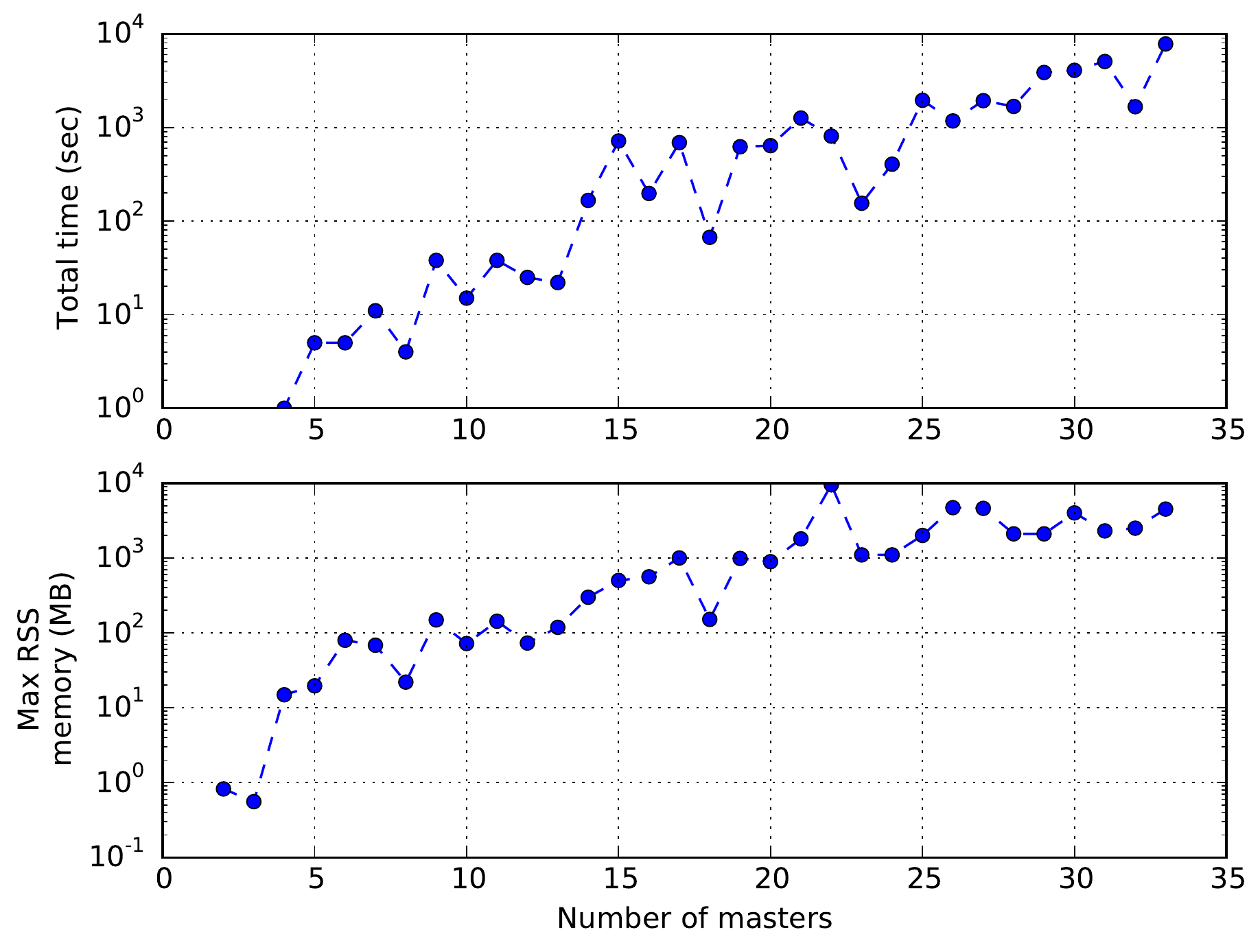}
\caption{BA, no reordering.}
\label{subfig:time_log}
\end{subfigure}
\caption{Selection of experimental measurements for the revised AMBA specification.}
\label{fig:amba}
\end{figure}

\paragraph{Revised specification}

The ARM processor Advanced Microcontroller Bus Architecture (AMBA) \cite{ARM99rev2} specifies a number of different bus protocols.
Among them, the Advanced High-performance (AHB) architecture has been studied extensively in the reactive synthesis literature \cite{Bloem07date,Bloem07cocv,Morgenstern10phd,Bloem12jcss,Schlaipfer12,Godhal13sttt,Bloem14amba}.

The AHB bus comprises of masters that need to communicate with slaves, and an arbiter that controls the bus and decides which master is given access to the bus.
The arbiter receives requests from the masters that desire to access the bus, and must respond in a weakly fair way.
In other words, every master that keeps uninterruptedly requesting the bus must eventually be granted access to it.
Note that the AMBA technical manual \cite{ARM99rev2} does not specify any fairness requirement, but instead leaves that decision to the designer.
For automated synthesis, weak fairness is one possible formalization that ensures servicing of all the masters.

In addition, a master can request that the access be locked.
In the ARM manual, the arbiter makes no promises as to whether a request for the lock will be granted.
If the arbiter does lock the access, then it guarantees to maintain the lock, until the request for locking is withdrawn by the master that currently owns the bus.
Note that the specification used here requires the arbiter to lock the bus, whenever requested by the master to be granted next.

A specification for the arbiter appeared in \cite{Bloem07date}, and is presented in detail in \cite{Bloem12jcss}.
Here, we expressed the specification of \cite{Bloem12jcss} in the proposed language, \cref{code:amba} on p.\pageref{code:amba}.
In doing so, some assumptions were weakened and assumption A1 modified, to improve the correspondence with the ARM technical manual, and reduce the number of environment variables (thus universal branching).
First, we describe the AHB specification, referring to \cref{code:amba}.
After that, we summarize the changes, and discuss the experiments.

The specification in \cref{code:amba} has both environment and system variables, as well as assumptions and guarantees.
The arbiter is the system, and the environment comprises of slaves and $N + 1$ masters.
An array \texttt{request} of bits represents the request of each individual master to be given bus ownership, for sending and receiving from a slave of interest.
Communication proceeds in bursts.
The bus owner selects which type of burst it desires, by setting the integer \texttt{burst}.
Three lengths of bursts are modeled: single time step (\texttt{SINGLE}), four consecutive time steps (\texttt{BURST4}), and undefined duration (\texttt{INCR}).
The currently addressed slave sets the bit \texttt{ready} (to true) to acknowledge that it has successfully received data for a burst.
While \texttt{ready} is false, the bus owner cannot change (G1,6), and a \texttt{BURST4} time step is not counted towards completion (G3).
For this reason, the slaves (environment) are required to recur setting \texttt{ready} to true (A2).
The master can also request that, when it is granted the bus, it should be locked.
Each master can do so by setting a signal.
Only two of these signals are modeled here, using the bit variables \texttt{grantee\_lockreq} and \texttt{master\_lockreq} (described below).

The arbiter works in primarily two phases, as introduced in \cite{Bloem12jcss}.
These phases are extraneous to the standard, and used only to aid in describing the specification.
Firstly, the arbiter decides to which master it will next grant the bus to.
The arbiter sets the bit \texttt{decide} to true during that period.
The decision is stored in the form of two variables, \texttt{grant} and \texttt{lockmemo}, which don't change while \texttt{decide} is false (G8).
The integer \texttt{grant} indicates the master that has been decided to receive bus ownership after the current owner.
For performance reasons, the arbiter can only grant the bus to a master that requested it (G10), with the exception of a default master (with index 0).

The bit \texttt{lockmemo} is set to the value of the environment bit \texttt{grantee\_lockreq} (G7).
The value \texttt{grantee\_lockreq'} represents whether the master \texttt{grant} had requested locked ownership.
In the original specification, an array $\mathit{lockreq}$ of $N$ environment bits is used (denoted by $\mathit{HLOCK}$ in \cite{Bloem12jcss}).
This increases significantly the variables with universal quantification.
Here, this array is abstracted by the bit \texttt{grantee\_lockreq}.
In implementation, the transducer input \texttt{grantee\_lockreq'} should be set equal to the lock request of master \texttt{grant} in the previous time step, i.e., $\mathit{grantee\_lockreq}' \triangleq (\previous \mathit{lockreq})[\mathit{grant}]$.
In \cite{Bloem12jcss}, some assumptions are expressed to constrain the array $\mathit{lockreq}$, i.e., when masters request locked ownership.
The assumptions can be weakened \cite{Filippidis15cds4-amba}, and by modifying assumption A1 (described below), the array $\mathit{lockreq}$ can be abstracted by the two bits \texttt{grantee\_lockreq} and \texttt{master\_lockeq}.

The arbiter promises to lock the bus, until the bus owner \texttt{master} interrupts requesting it.
The owner indicates its lock request by the value \texttt{master\_lockreq'}.
In implementation, the input value \texttt{master\_lockreq'} should be set equal to the lock request of \texttt{master}, i.e., $\mathit{master\_lockreq}' \triangleq \mathit{lockreq}[\mathit{master}]$.

In the second phase, the master changes the bus owner, by updating the integer \texttt{master} to \texttt{grant} (G4,5).
If the grantee had requested a lock, via \texttt{grantee\_lockreq}, then that request is propagated to the bit \texttt{lock} (G4,5).
With the bit \texttt{lock}, the arbiter indicates that \texttt{master} has been given locked access to the bus.

To be weakly fair, each master that keeps uninterruptedly requesting the bus should be granted ownership.
This requirement is described as a B\"{u}chi automaton (G9).

The assumption A1 of \cite{Bloem12jcss} requires that for locked undefined-length bursts, the masters eventually withdraw their request to access the bus.
This assumption is not explicit in the ARM standard, so we modify it, by requiring that masters withdraw only their request for the lock, {\em not} for bus access.
This is described as the B\"{u}chi automaton \texttt{withdraw\_lock} that constrains the environment.
The arbiter grants \texttt{master} locked access by setting the bit \texttt{lock} to true.
If \texttt{lock} is false, then the master (environment) remains in the outer loop, at the \texttt{else}.
If \texttt{lock} becomes true, then the automaton enters the inner loop.
In order for the automaton \texttt{withdraw\_lock} to exit the inner loop, the environment must set \texttt{master\_lockreq'} to false.
This obliges the owner \texttt{master} to eventually stop requesting locked ownership.

For a \texttt{SINGLE} burst, the burst is completed at the next time step that \texttt{ready} is true, so the arbiter does not need to lock the bus (since the owner remains unchanged while \texttt{ready} is false).
For a \texttt{BURST4} burst, the arbiter locks the bus for a predefined length of four successful beats (G3).
This requirement is described by the safety automaton \texttt{count\_bursts}.
Note that assumption A1 is not needed for this case.
For a \texttt{INCR} burst, the duration is unspecified a priori.
While the owner \texttt{master} continuously requests locked access (with \texttt{master\_lockreq}), the arbiter cannot change the bus owner (G2).
This is described by the safety automaton \texttt{maintain\_lock}.
When the arbiter grants locked access to the bus for a burst of undefined duration, then the guard \texttt{lock \&\& start \&\& (burst == INCR)} is true.
The process \texttt{maintain\_lock} enters the inner loop, and remains there until \texttt{master\_lockreq} becomes true.
This is where assumption A1 is required, to ensure that the owner will eventually stop requesting the lock.
The arbiter can then exit the inner loop of \texttt{maintain\_lock}.
Then, the arbiter can wait outside (\texttt{start} is false throughout the burst), until the addressed slave sets \texttt{ready} to true, signifying the successful completion of that burst, and allowing the arbiter to set \texttt{start} and change the bus owner, if needed.

Some properties not in GR(1) are translated to deterministic B\"{u}chi automata in \cite{Bloem12jcss}.
The resulting formulae are much less readable, and not easy to modify and experiment with.
Above, we specified these properties directly as processes, with progress states where needed.

\begin{figure}[t]
\centering
\begin{subfigure}{0.49\textwidth}
\includegraphics[width=\textwidth]{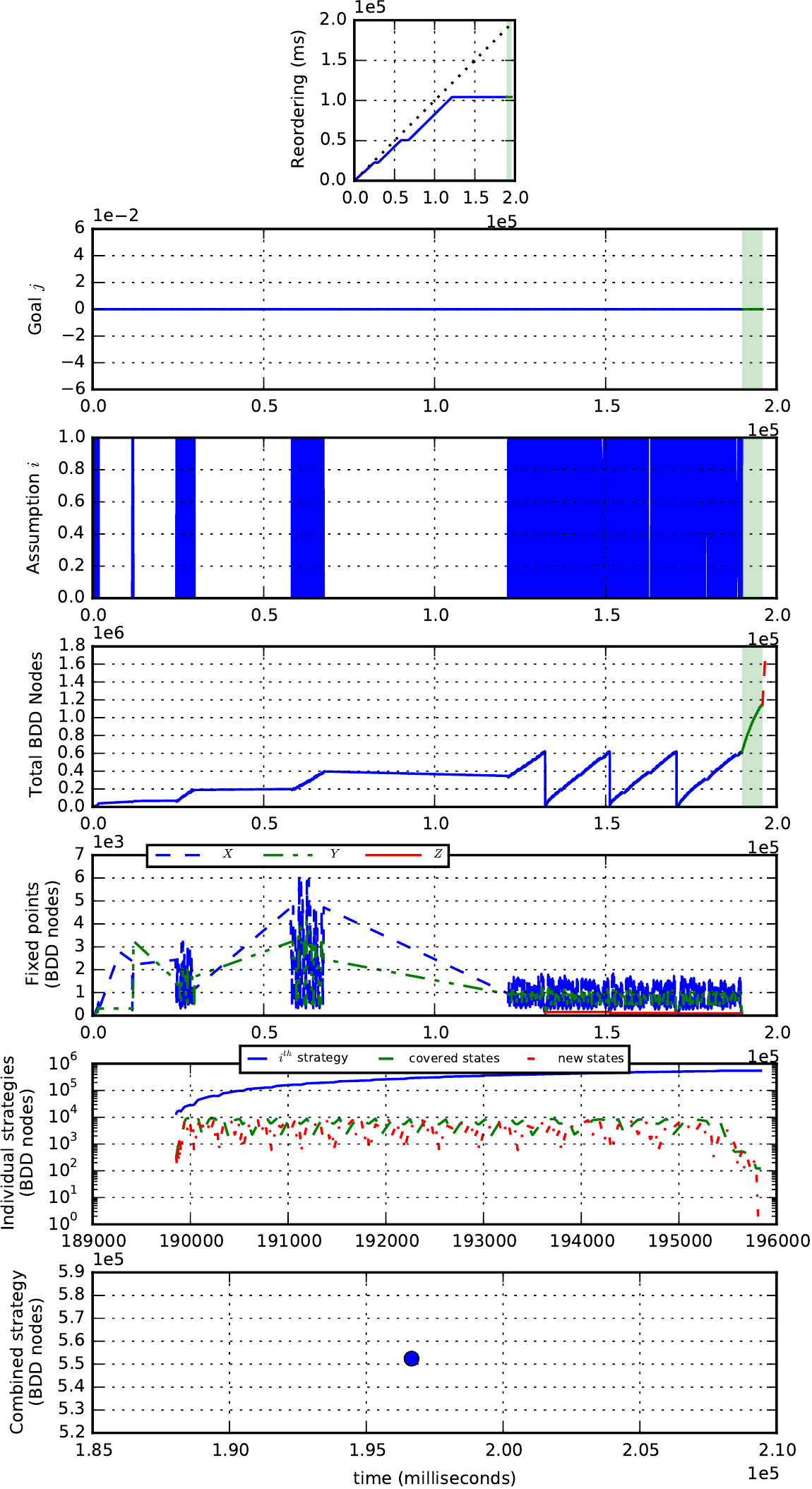}
\caption{BA, no reordering, $N = 16$.}
\label{subfig:details_16_ba_mine_no}
\end{subfigure}
\begin{subfigure}{0.49\textwidth}
\includegraphics[width=\textwidth]{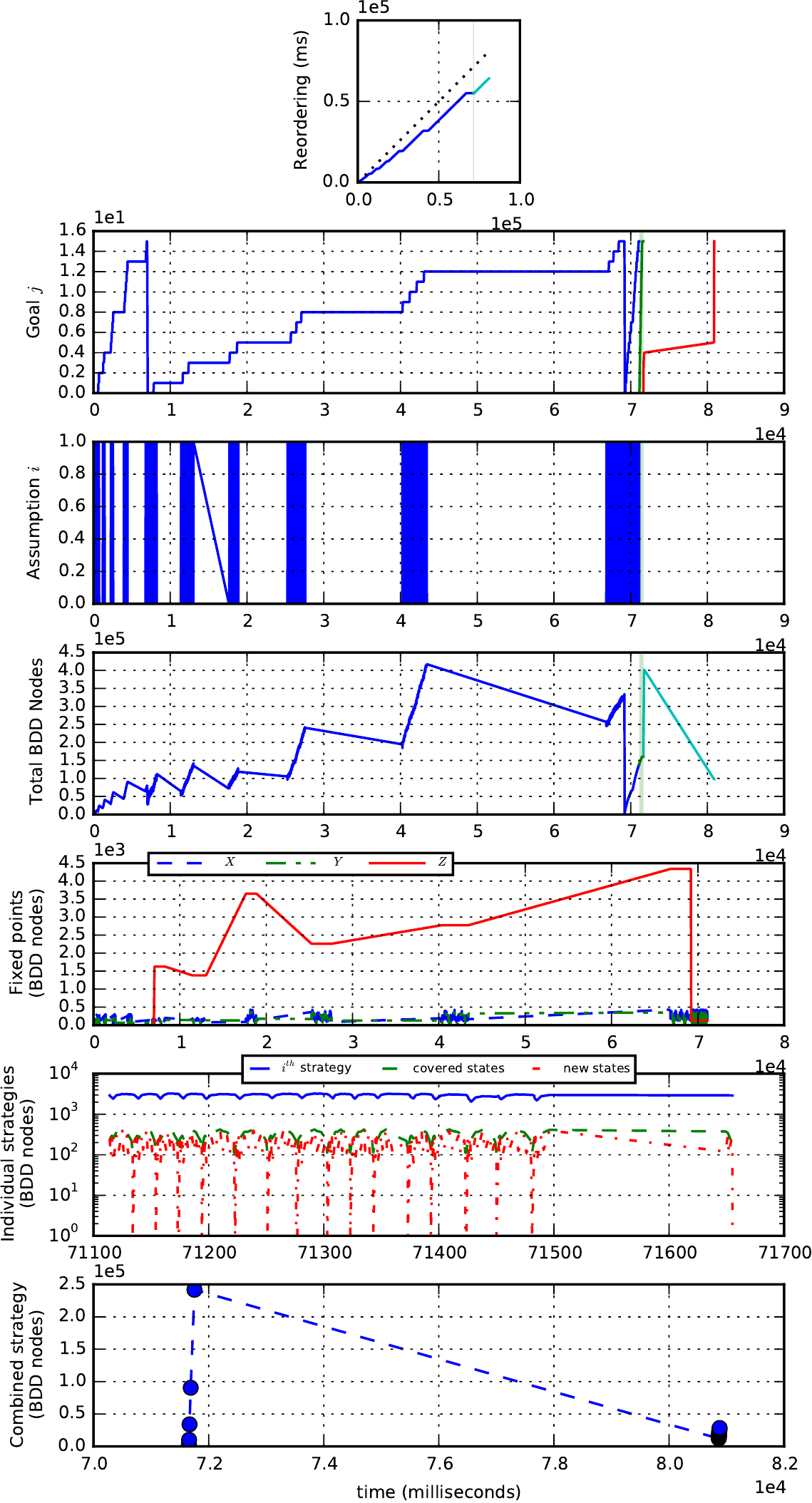}
\caption{Conjunction, with reordering, $N = 16$.}
\end{subfigure}
\caption{
	Measurements during phases of: (1) realizability, (2) sub-strategy, and (3) combined strategy construction.
	The top 4 plots are over all phases, the fixpoints over (1), the individual strategies over (2), and the bottom plot over (3).
	The revised specification is used.}
\label{fig:amba-details}
\end{figure}

\paragraph{Observations}

By encoding \texttt{master} and \texttt{grant} as integers, and abstracting the array $\mathit{lockreq}$ by the two variables \texttt{master\_lockreq} and \texttt{grantee\_lockreq}, the synthesis time was reduced significantly (by a factor of 100 \cite{Filippidis15cds4-amba}), but are not sufficient to prevent the synthesized strategies from blowing up.
By also merging the $N$ weak fairness guarantees
$\bigwedge_{i = 0}^{N - 1} \always \eventually ( \mathit{request}[i] \rightarrow \mathit{master} = i)$
into the B\"{u}chi automaton (BA) \texttt{weak\_fairness} with one accepting state, we were able to prevent the strategies from blowing up, and synthesize up to 33 masters, \cref{subfig:time_log}.
The synthesis time for 16 masters is in the order of 5 minutes, and peak memory consumption less than 1GB.
To our knowledge, in previous works, the maximal number of masters has been 16, the strategies were blowing up, and the runtimes were significantly longer (21 hours for 12 masters in \cite{Bloem12jcss}, and more than an hour in \cite{Godhal13sttt} for 16 masters).

\paragraph{Measurements}
\label{sec:measurements}

To identify what caused this difference, we conducted experiments for 8 different combinations: original vs revised spec, conjunction vs BA, reordering during strategy construction enabled/disabled, \cref{tab:results-overview}.
We collected detailed measurements with instrumentation that we inserted into \textsc{Slugs}, available at \cite{slugsfork}.
Some of these are shown in \cref{fig:amba-details}, and the complete set can be found in the technical report \cite{Filippidis15cds4-amba} (the language is described in \cite{Filippidis15cds3-synt}).
The experiments were run on an Intel(R) Xeon\textsuperscript{\textregistered} X5550 core, with 27 GB RAM, running Ubuntu 14.04.1.
The maximal memory limit of \textsc{Cudd} \cite{Somenzi12cudd} was set to 16 GB.

We found that lack of dynamic BDD reordering during construction of the strategy was the reason for poor performance of conjoined liveness goals, as opposed to a single BA.
The implementation of the GR(1) synthesis algorithm in \textsc{Slugs} has three phases:
\begin{enumerate}
	\item Computing the winning region, while memoizing the iterates of the fixpoint iteration, as BDDs.
	\item Construction of individual strategies, one for each recurrence goal.
	\item Combination of the individual strategies into a single transducer, which iterates through them.
\end{enumerate}
In \textsc{Slugs}, variable reordering \cite{Rudell1993iccad} is enabled during the first two phases, but disabled in the last one.
If the recurrence goals are conjoined into a formula of the form $\bigwedge \always \eventually$, then the memory needed for synthesis blows up \cref{subfig:comparison_ba_mine_no_conj_mine_no}, for both the original and revised specifications.
Using a BA, the revised specification scales without blowup.

If reordering is enabled during the last phase (combined transducer construction), then the specification with conjunction can be synthesized without blowup.
With a BA, enabling reordering in the last phase has a mildly negative effect, because it can trigger unnecessary reordering.
We used the group sifting algorithm \cite{Panda95iccad,Rudell1993iccad} for reordering.

Enabling dynamic BDD variable reordering  is necessary to prevent the blowup.
The conjunction with reordering enabled in phase 3 outperforms the BA with reordering turned off in phase 3.
This is a consequence mainly of the fact that the BA chains the goals inside the state space, leading to deeper fixpoint iterations, and has slightly larger state space, due to the nodes of the automaton \texttt{maintain\_lock}.

Reordering typically accounts for most of the runtime (top plot in \cref{subfig:details_16_ba_mine_no}).
The second plot shows the currently pursued goal during realizability, and later the sub-strategy under construction, and the sub-strategy being combined in the final strategy.
Each drop in total BDD nodes (“teeth” in 4th plot) corresponds to an outer fixpoint iteration.
The first outer iteration takes the most time, due to reordering.
Later iterations construct subsets, for which the obtained order remains suitable.
The highlighted period corresponds to the construction of individual strategies.
Plots for the other experiments can be found in \cite{Filippidis15cds4-amba}.

\begin{table}
\caption{Overview of results.}
\label{tab:results-overview}
\centering
\begin{tabular}{cccc}
\toprule
& \multirow{2}{1.5cm}{Strategy reordering} & \multicolumn{2}{c}{Specification}\\
& & original & revised\\
\midrule
\multirow{2}{*}{Conjunction of fairness} & with & slow & fast\\
& w/o & memory blowup & memory blowup\\
\midrule
\multirow{2}{*}{B\"{u}chi automaton} & with & very slow & ok (slower)\\
& w/o & slow & ok\\
\bottomrule
\end{tabular}
\end{table}

\paragraph{Conclusions}

The major effect of reordering in the final phase of strategy construction can be understood as follows.
Using a BA reduces the goals to only one, so no disjunction of individual sub-strategies is needed \cite{Piterman06}.
Also, this encoding shifts the transducer memory (a counter of liveness goals), from the strategy construction, to the realizability phase (attractor computations).
This slightly increases the state space.
Nonetheless, this symbolic encoding allows the variable ordering more time to gradually adjust to the represented sets.

In contrast, by conjoining liveness goals, the variable order is oblivious during realizability checking that the sub-strategies will be disjoined at the end.
The disjunction of strategies acts as a shock wave, disruptive to how far from optimal the obtained variable order is.
If, by that phase, reordering has been disabled, then this effect causes exponential blowup.

Overall, the proposed language made experimentation easier and revisions faster, helping to study variants of the specification.
It can be used to explore the sensitivity of a specification, in the following way.
A formula, e.g., requiring weak fairness, can be temporarily replaced with a process that is one possible refinement of that formula, potentially simplified.
In the AMBA example, one can fix a round robin schedule for selecting the next grantee (temporarily dropping G10).
This is reminiscent of the manual implementation \cite{Bloem12jcss}.
By doing so, it can be evaluated whether the synthesizer finds it difficult to pick requestors only, or whether some other factor is more important, either another part of the specification, or some external factor.
For the AMBA problem, such a change resulted in replacing recurrence formulae with a BA, and led to identifying the need for strategy reordering to avoid memory blowup.
Therefore, we believe that it can prove useful in exploring the sensitivity of specifications, to help the specifier direct their attention to improve those parts of the specification that impact the most synthesis performance.

\begin{lstlisting}[basicstyle=\ttfamily\small,xleftmargin=3em,multicols=2,caption={AMBA AHB specification in the proposed language.},label={code:amba}]
#define N 2  /* N + 1 masters */
#define SINGLE 0
#define BURST4 1
#define INCR 2
/* variables of masters and slaves
A4: initial condition */
free env bool ready = false;
free env int(0, 2) burst;
free env bool request[N + 1] =false;
free env bool grantee_lockreq=false;
free env bool master_lockreq =false;
/* arbiter variables */
/* G11: sys initial condition */
free bool start = true;
free bool decide = true;
free bool lock = false;
free bool lockmemo;
free int(0, N) master = 0;
free int(0, N) grant;
/* A2: slaves must progress with receiving data */
assume ltl { []<> ready }
/* A3: dropped, weakening the assumptions */
/* A1: */
assume env proctype withdraw_lock(){
    progress:
    do
    :: lock;
        do
        :: ! master_lockreq'; break
        :: true /* wait */
        od
    :: else
    od
}
assert ltl {
[](
/* G1: new access starts only when slave is ready */
(start' -> ready)
/* G4,5 */
&& (ready -> ((master' == grant) && (lock' <-> lockmemo')))
/* G6 */
&& (! start' -> (
    (master' == master) &&
    (lock' <-> lock)))
/* G7: remember if lock requested */
&& ((--X decide) -> (lockmemo' <-> grantee_lockreq'))
/* G8 */
&& (! decide -> (grant' == grant))
&& ((! --X decide) -> (lockmemo' <-> lockmemo))
/* G10: grant only to requestors */
&& ((grant' == grant) || (grant' == 0) || request[grant'])
)
}
sync{ /* synchronous product */
/* G9: weak fairness */
assert proctype weak_fairness(){
    int(0, N) count;
    do
    :: (! request[count] || (master == count));
        if
        :: (count < N) && (count' == count + 1)
        :: (count == N) && (count' == 0);
            progress: skip
        fi
    :: else
    od
}
/* G2: lock until no lock req */
assert sys proctype maintain_lock(){
    do
    :: (lock && start && (burst == INCR));
        do
        :: (! start && ! master_lockreq'); break
        :: ! start
        od
    :: else
    od
}
/* G3: for a BURST4 access, count the "ready" time steps. */
assert sys proctype count_burst(){
    int(0, 3) count;
    do
    :: (start && lock &&
        (burst == BURST4) &&
        (!ready || (count' == 1)) &&
        (ready || (count' == 0)) );
        do
        :: (! start && ! ready)
        :: (! start && ready && (count < 3) &&
            (count' == count + 1))
        :: (! start && ready && (count >= 3)); break
        od
    :: else
    od
}
}
\end{lstlisting}

\section{Relevant work}
\label{sec:relevant-work}

Our approach has common elements with program repair \cite{Jobstmann05cav}, program sketching \cite{Lezama08phd}, and syntax-guided synthesis \cite{Alur2013fmcad}.
Program repair aims at modifying an existing program in a conventional programming language.
Syntax-guided synthesis uses a grammar to “slice” the admissible search space of terminating programs.
Here, we are interested in reactive programs.
Similarly, program sketching uses templates to restrict the search space and give hints to the synthesizer for obtaining a complete program.
In \cite{Beyene2014popl}, the authors propose another constraint-based approach to games, but start directly from logic formulae.

TLA \cite{Lamport94tpls,Lamport02} subsumes our proposed language, since it includes quantification, but is intended as a theorem proving activity, is declarative, and is aimed at verification.
Nonetheless, one can view the proposed translation as from open-\textsc{Promela} to TLA.
\textsc{SMV} is a declarative language \cite{Cavada10nusmv}, and \textsc{Jtlv} \cite{Pnueli10cav} an \textsc{SMV}-like language for synthesis specifications, but with no imperative constructs.
\textsc{AspectLTL} is a further declarative extension for aspect-oriented programming \cite{Maoz11aosd}.

\textsc{RPromela} is an extension of \textsc{Promela} that adds synchronous-reactive constructs (not in the sense of reactive synthesis) that include synchronous products and channels called ports \cite{Najm96spin,Najm96tacas}.
Its semantics are defined in terms of {\em stable states}, where the synchronous product blocks, waiting for message reception from its global ports.
\textsc{RPromela} does not address modeling of the environment, nor declarative elements.
Besides, synchronous-reactive languages like \textsc{Esterel}, \textsc{Quartz} (imperative textual), \textsc{Statecharts}, \textsc{Argos}, \textsc{SyncCharts} (imperative graphical), \textsc{Lustre}, and \textsc{Lucid Synchrone} (declarative textual) and \textsc{Signal} (declarative graphical) are by definition {\em deterministic} languages intended for direct design of transducers \cite{Gamatie09,Halbwachs92,Jourdan94iccl}.
In synthesis, non-determinism is an essential feature of the specification.

Our approach has common elements with {\em constraint imperative programming} (CIP), introduced with the experimental language \textsc{Kaleidoscope} \cite{FreemanBenson90ecoop,FreemanBenson92ecoop,FreemanBenson92iccl,Lopez94oopsla}, one of the first attempts to integrate the imperative and declarative constraint programming paradigms.
An observation from \cite{FreemanBenson90ecoop}, which applies also here, is that specifiers need to express two types of relations: long-lived (best described declaratively), and sequencing relations (more naturally expressed in an imperative style).
However, CIP does {\em not} ensure correct reactivity, because the constraints are solved online.
Constraints are a related approach that uses constraints for indirect assignment to imperative variables is \cite{Lamport85popl}.

The translation from \textsc{Promela} to declarative formalisms has been considered in \cite{Baldamus01spin,Yong12spin,Ciesinski08spin} and decision diagrams in \cite{Beaudenon10sttt}.
These translations aim at verification, do not have LTL as target language, and either have limited support for atomicity \cite{Ciesinski08spin}, no details \cite{Baldamus01spin}, or programs graphs semantics that do not match \textsc{Promela} \cite{Yong12spin}.

\section{Conclusions}
\label{sec:conclusions}

We have presented a language for reactive synthesis that combines declarative and imperative elements to allow using the most suitable paradigm for each requirement, to write readable specifications.
By expressing the AMBA specification in a multi-paradigm language, it became easier to experiment and transform it into one that led to efficient synthesis that improved previous results by two orders of magnitude.
Besides the AMBA specification, other examples can be found in the code repository \cite{Filippidis15github-openpromela}.

\paragraph{Acknowledgments}

The authors would like to thank Scott Livingston for providing helpful feedback.
This work was supported by STARnet, a Semiconductor Research Corporation program, sponsored by MARCO and DARPA. 
The first author was partially supported by a graduate research fellowship from the Jet Propulsion Laboratory, over the summer of 2014.

\bibliographystyle{eptcs}
\bibliography{literature}

\end{document}